\RequirePackage{lineno}
\documentclass[aps,prl,twocolumn,amsmath,amssymb,superscriptaddress,showpacs]{revtex4}

\usepackage{graphicx}
\usepackage{dcolumn}
\usepackage{bm}
\usepackage[abs]{overpic} 
\usepackage{feynmf}

\preprint{draftv21.0} 

   \newcommand{\Bd}{\ensuremath{B^{0}}}                                                                                                   %
   \newcommand{\Bs}{\ensuremath{B_{s}^{0}}}                                                                                               %
   \newcommand{\Bdpipi}{\ensuremath{\Bd\rightarrow\pi^{+}\pi^{-}}}                                                                       %
   \newcommand{\BdKpi}{\ensuremath{\Bd\rightarrow K^{+}\pi^{-}}}                                                                        %
   \newcommand{\BdKK}{\ensuremath{\Bd\rightarrow K^{+}K^{-}}}                                                                           %
   \newcommand{\BsKK}{\ensuremath{\Bs\rightarrow K^{+}K^{-}}}                                                                           %
   \newcommand{\BsKpi}{\ensuremath{\Bs\rightarrow K^{-}\pi^{+}}}                                                                        %
   \newcommand{\Bspipi}{\ensuremath{\Bs\rightarrow\pi^{+}\pi^{-}}}

   \newcommand{\DKpi}{\ensuremath{D^0\rightarrow K^-\pi^+}}      
   \newcommand{\Dpipi}{\ensuremath{D^0\rightarrow \pi^+\pi^-}}             
  \newcommand{\Lb}{\ensuremath{\Lambda^0_b}}           
   
   \newcommand{\BspipisuBdKpidef}{\ensuremath{\frac{f_s}{f_d}\frac{\BR(\Bs\rightarrow \pi^+\pi^-)}{\BR(\Bd\rightarrow K^+\pi^-)}}}
   \newcommand{\BdKKsuBdKpidef}{\ensuremath{\frac{\BR(\Bd\rightarrow K^+K^-)}{\BR(\Bd\rightarrow K^+\pi^-)}}}

   \newcommand{\BR}{\ensuremath{\mathcal B}}                                                                                   %
   \newcommand{\dedx}{\ensuremath{\mathit{dE/dx}}}                                                                              %
   \newcommand{\micron}{\ensuremath{\mu\mathrm{m}}}
   \newcommand{\pgev}{\ensuremath{\mathrm{GeV}/c}}
   \newcommand{\ptot}{\ensuremath{p_{\mathrm{tot}}}}   
   \newcommand{\massgev}{\ensuremath{\mathrm{GeV}/c^{2}}}

\newcommand{\lumifb}{ fb$^{-1}$}
\newcommand{\like}{\ensuremath{\mathcal{L}}}

\newcommand{\mpipi}{\ensuremath{m_{\pi^+\pi^-}}}

\newcommand{\stat}{\ensuremath{\mathrm{~(stat)}}}		
\newcommand{\syst}{\ensuremath{\mathrm{~(syst)}}}		

\newcommand{\Lxy}{\ensuremath{L_{T}}}			
\begin{document}
\title{Evidence for the charmless annihilation decay mode \Bspipi}
\affiliation{Institute of Physics, Academia Sinica, Taipei, Taiwan 11529, Republic of China}
\affiliation{Argonne National Laboratory, Argonne, Illinois 60439, USA}
\affiliation{University of Athens, 157 71 Athens, Greece}
\affiliation{Institut de Fisica d'Altes Energies, ICREA, Universitat Autonoma de Barcelona, E-08193, Bellaterra (Barcelona), Spain}
\affiliation{Baylor University, Waco, Texas 76798, USA}
\affiliation{Istituto Nazionale di Fisica Nucleare Bologna, $^{cc}$University of Bologna, I-40127 Bologna, Italy}
\affiliation{University of California, Davis, Davis, California 95616, USA}
\affiliation{University of California, Los Angeles, Los Angeles, California 90024, USA}
\affiliation{Instituto de Fisica de Cantabria, CSIC-University of Cantabria, 39005 Santander, Spain}
\affiliation{Carnegie Mellon University, Pittsburgh, Pennsylvania 15213, USA}
\affiliation{Enrico Fermi Institute, University of Chicago, Chicago, Illinois 60637, USA}
\affiliation{Comenius University, 842 48 Bratislava, Slovakia; Institute of Experimental Physics, 040 01 Kosice, Slovakia}
\affiliation{Joint Institute for Nuclear Research, RU-141980 Dubna, Russia}
\affiliation{Duke University, Durham, North Carolina 27708, USA}
\affiliation{Fermi National Accelerator Laboratory, Batavia, Illinois 60510, USA}
\affiliation{University of Florida, Gainesville, Florida 32611, USA}
\affiliation{Laboratori Nazionali di Frascati, Istituto Nazionale di Fisica Nucleare, I-00044 Frascati, Italy}
\affiliation{University of Geneva, CH-1211 Geneva 4, Switzerland}
\affiliation{Glasgow University, Glasgow G12 8QQ, United Kingdom}
\affiliation{Harvard University, Cambridge, Massachusetts 02138, USA}
\affiliation{Division of High Energy Physics, Department of Physics, University of Helsinki and Helsinki Institute of Physics, FIN-00014, Helsinki, Finland}
\affiliation{University of Illinois, Urbana, Illinois 61801, USA}
\affiliation{The Johns Hopkins University, Baltimore, Maryland 21218, USA}
\affiliation{Institut f\"{u}r Experimentelle Kernphysik, Karlsruhe Institute of Technology, D-76131 Karlsruhe, Germany}
\affiliation{Center for High Energy Physics: Kyungpook National University, Daegu 702-701, Korea; Seoul National University, Seoul 151-742, Korea; Sungkyunkwan University, Suwon 440-746, Korea; Korea Institute of Science and Technology Information, Daejeon 305-806, Korea; Chonnam National University, Gwangju 500-757, Korea; Chonbuk National University, Jeonju 561-756, Korea}
\affiliation{Ernest Orlando Lawrence Berkeley National Laboratory, Berkeley, California 94720, USA}
\affiliation{University of Liverpool, Liverpool L69 7ZE, United Kingdom}
\affiliation{University College London, London WC1E 6BT, United Kingdom}
\affiliation{Centro de Investigaciones Energeticas Medioambientales y Tecnologicas, E-28040 Madrid, Spain}
\affiliation{Massachusetts Institute of Technology, Cambridge, Massachusetts 02139, USA}
\affiliation{Institute of Particle Physics: McGill University, Montr\'{e}al, Qu\'{e}bec, Canada H3A~2T8; Simon Fraser University, Burnaby, British Columbia, Canada V5A~1S6; University of Toronto, Toronto, Ontario, Canada M5S~1A7; and TRIUMF, Vancouver, British Columbia, Canada V6T~2A3}
\affiliation{University of Michigan, Ann Arbor, Michigan 48109, USA}
\affiliation{Michigan State University, East Lansing, Michigan 48824, USA}
\affiliation{Institution for Theoretical and Experimental Physics, ITEP, Moscow 117259, Russia}
\affiliation{University of New Mexico, Albuquerque, New Mexico 87131, USA}
\affiliation{The Ohio State University, Columbus, Ohio 43210, USA}
\affiliation{Okayama University, Okayama 700-8530, Japan}
\affiliation{Osaka City University, Osaka 588, Japan}
\affiliation{University of Oxford, Oxford OX1 3RH, United Kingdom}
\affiliation{Istituto Nazionale di Fisica Nucleare, Sezione di Padova-Trento, $^{dd}$University of Padova, I-35131 Padova, Italy}
\affiliation{University of Pennsylvania, Philadelphia, Pennsylvania 19104, USA}
\affiliation{Istituto Nazionale di Fisica Nucleare Pisa, $^{ee}$University of Pisa, $^{ff}$University of Siena and $^{gg}$Scuola Normale Superiore, I-56127 Pisa, Italy}
\affiliation{University of Pittsburgh, Pittsburgh, Pennsylvania 15260, USA}
\affiliation{Purdue University, West Lafayette, Indiana 47907, USA}
\affiliation{University of Rochester, Rochester, New York 14627, USA}
\affiliation{The Rockefeller University, New York, New York 10065, USA}
\affiliation{Istituto Nazionale di Fisica Nucleare, Sezione di Roma 1, $^{hh}$Sapienza Universit\`{a} di Roma, I-00185 Roma, Italy}
\affiliation{Rutgers University, Piscataway, New Jersey 08855, USA}
\affiliation{Texas A\&M University, College Station, Texas 77843, USA}
\affiliation{Istituto Nazionale di Fisica Nucleare Trieste/Udine, I-34100 Trieste, $^{ii}$University of Udine, I-33100 Udine, Italy}
\affiliation{University of Tsukuba, Tsukuba, Ibaraki 305, Japan}
\affiliation{Tufts University, Medford, Massachusetts 02155, USA}
\affiliation{University of Virginia, Charlottesville, Virginia 22906, USA}
\affiliation{Waseda University, Tokyo 169, Japan}
\affiliation{Wayne State University, Detroit, Michigan 48201, USA}
\affiliation{University of Wisconsin, Madison, Wisconsin 53706, USA}
\affiliation{Yale University, New Haven, Connecticut 06520, USA}

\author{T.~Aaltonen}
\affiliation{Division of High Energy Physics, Department of Physics, University of Helsinki and Helsinki Institute of Physics, FIN-00014, Helsinki, Finland}
\author{B.~\'{A}lvarez~Gonz\'{a}lez$^x$}
\affiliation{Instituto de Fisica de Cantabria, CSIC-University of Cantabria, 39005 Santander, Spain}
\author{S.~Amerio}
\affiliation{Istituto Nazionale di Fisica Nucleare, Sezione di Padova-Trento, $^{dd}$University of Padova, I-35131 Padova, Italy}
\author{D.~Amidei}
\affiliation{University of Michigan, Ann Arbor, Michigan 48109, USA}
\author{A.~Anastassov$^v$}
\affiliation{Fermi National Accelerator Laboratory, Batavia, Illinois 60510, USA}
\author{A.~Annovi}
\affiliation{Laboratori Nazionali di Frascati, Istituto Nazionale di Fisica Nucleare, I-00044 Frascati, Italy}
\author{J.~Antos}
\affiliation{Comenius University, 842 48 Bratislava, Slovakia; Institute of Experimental Physics, 040 01 Kosice, Slovakia}
\author{G.~Apollinari}
\affiliation{Fermi National Accelerator Laboratory, Batavia, Illinois 60510, USA}
\author{J.A.~Appel}
\affiliation{Fermi National Accelerator Laboratory, Batavia, Illinois 60510, USA}
\author{T.~Arisawa}
\affiliation{Waseda University, Tokyo 169, Japan}
% ...

\collaboration{CDF Collaboration\footnote{With visitors from
$^a$Istituto Nazionale di Fisica Nucleare, Sezione di Cagliari, 09042 Monserrato (Cagliari), Italy,
$^b$University of CA Irvine, Irvine, CA 92697, USA,
$^c$University of CA Santa Barbara, Santa Barbara, CA 93106, USA,
$^d$University of CA Santa Cruz, Santa Cruz, CA 95064, USA,
$^e$CERN, CH-1211 Geneva, Switzerland,
$^f$Cornell University, Ithaca, NY 14853, USA,
$^g$University of Cyprus, Nicosia CY-1678, Cyprus,
$^h$Office of Science, U.S. Department of Energy, Washington, DC 20585, USA,
$^i$University College Dublin, Dublin 4, Ireland,
$^j$ETH, 8092 Zurich, Switzerland,
$^k$University of Fukui, Fukui City, Fukui Prefecture, Japan 910-0017,
$^l$Universidad Iberoamericana, Mexico D.F., Mexico,
$^m$University of Iowa, Iowa City, IA 52242, USA,
$^n$Kinki University, Higashi-Osaka City, Japan 577-8502,
$^o$Kansas State University, Manhattan, KS 66506, USA,
$^p$University of Manchester, Manchester M13 9PL, United Kingdom,
$^q$Queen Mary, University of London, London, E1 4NS, United Kingdom,
$^r$University of Melbourne, Victoria 3010, Australia,
$^s$Muons, Inc., Batavia, IL 60510, USA,
$^t$Nagasaki Institute of Applied Science, Nagasaki, Japan,
$^u$National Research Nuclear University, Moscow, Russia,
$^v$Northwestern University, Evanston, IL 60208, USA,
$^w$University of Notre Dame, Notre Dame, IN 46556, USA,
$^x$Universidad de Oviedo, E-33007 Oviedo, Spain,
$^y$CNRS-IN2P3, Paris, F-75205 France,
$^z$Texas Tech University, Lubbock, TX 79609, USA,
$^{aa}$Universidad Tecnica Federico Santa Maria, 110v Valparaiso, Chile,
$^{bb}$Yarmouk University, Irbid 211-63, Jordan.
}}
\noaffiliation

\date{November 1, 2011}
\begin{abstract}
We search for annihilation decay modes of neutral $b$~mesons into
pairs of charmless charged hadrons with the upgraded Collider Detector at the Fermilab Tevatron.
Using a data sample corresponding to \mbox{6\lumifb} of integrated luminosity, 
we obtain the first evidence for the \Bspipi\ decay, with a significance of 
$3.7\sigma$, and a measured branching ratio $\BR(\Bspipi)= (0.57 \pm 0.15\stat \pm 0.10\syst)\times 10^{-6}$. 
A search for the \BdKK\ mode in the same sample yields a significance of $2.0\sigma$, 
and a central value estimate $\BR(\BdKK)= (0.23 \pm 0.10\stat \pm 0.10\syst)\times 10^{-6}$.
\end{abstract}

\maketitle


Our understanding of the dynamics of hadrons containing heavy quarks
has made great progress in recent years. The development of effective theories 
has allowed increasingly accurate predictions for the partial decay widths of such hadrons.  
An ability to make accurate predictions for these processes is not only important in itself, 
but is a tool to uncover possible additional contributions due to interactions beyond the standard model.
In spite of the general progress of the field, a specific class of decay amplitudes (annihilation topologies) 
has resisted attempts at quantitative prediction up to the present, and is often simply neglected in 
calculations. Predictions for these amplitudes vary greatly between approaches, and even within the same approach. 
Estimates based on the QCD factorization (QCDF)
approach are affected by significant uncertainties, due to end-point singularities~\cite{Beneke:2003zv,Cheng:2009mu}. 
More recent perturbative QCD calculations (pQCD) provide more precise predictions, but they tend to be significantly larger 
than the predictions coming from QCDF~\cite{Ali:2007ff,Li:2004ep}.
No calculations are yet available within the soft collinear effective theory (SCET)~\cite{SCET:2000}. 
%
\begin{figure}[htb]
\includegraphics[scale=0.3]{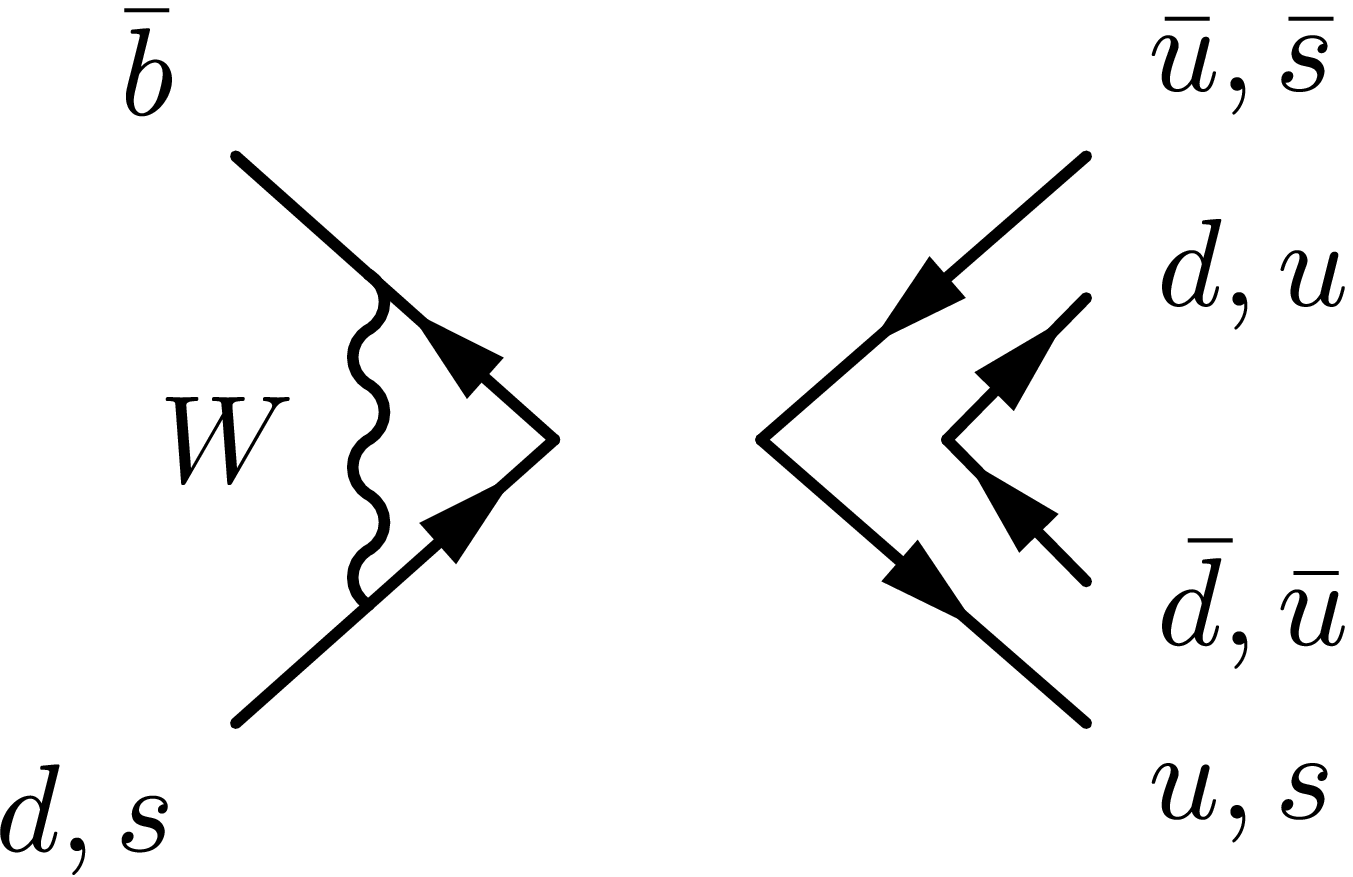}
\includegraphics[scale=0.3]{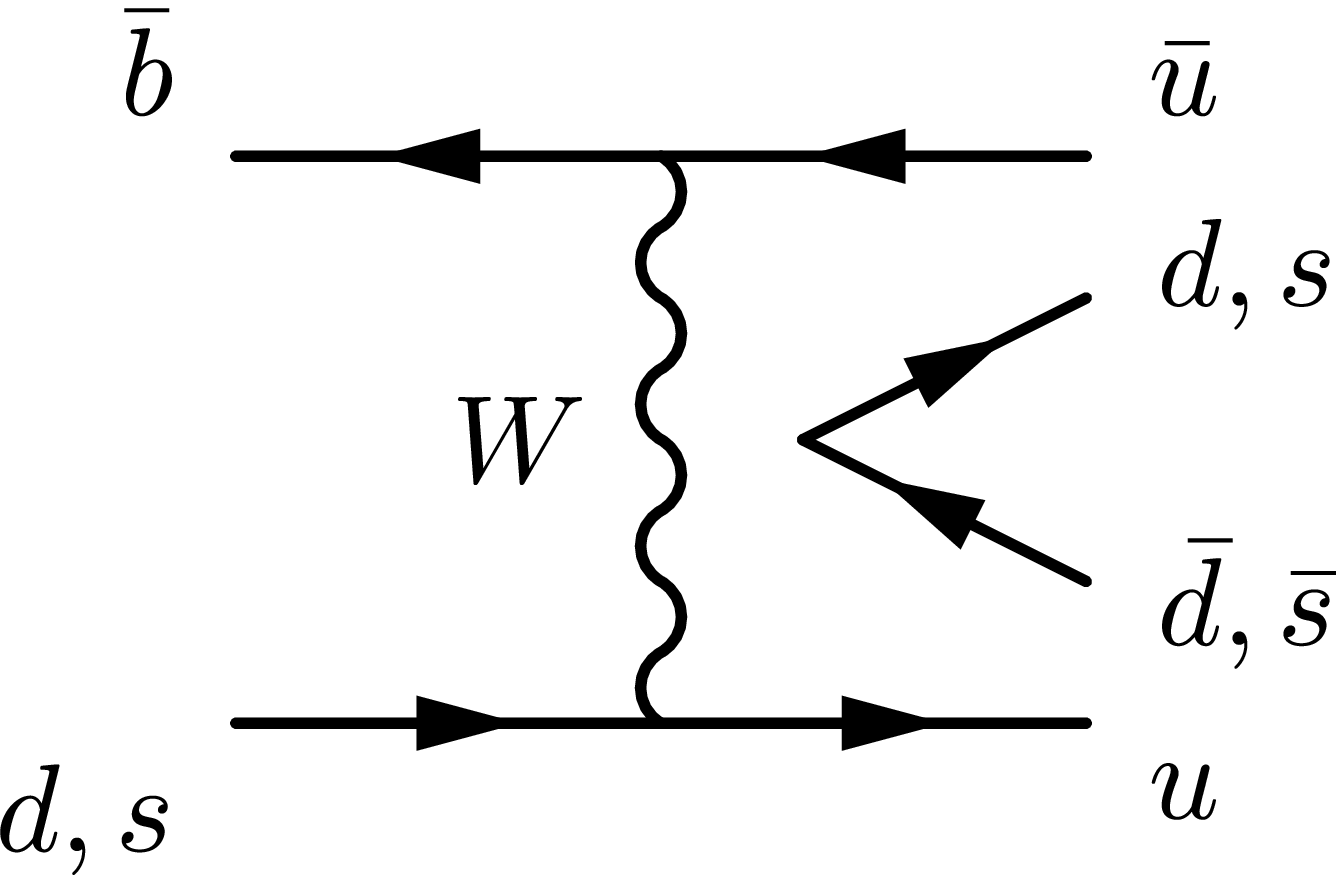}
\caption[$W$-exchange and penguin-annihilation diagrams]{$\mathit{PA}$ (left panel) and $E$ (right panel) diagrams 
contributing to \BdKK\ and \Bspipi\ decays.}
\label{fig:diagrams}
\end{figure}
%
The lack of knowledge of the size of annihilation-type amplitudes
introduces irreducible uncertainties in the predictions 
for several decays of great interest in the search for new physics effects, 
such as \Bdpipi\ and \BsKK~\cite{Soni:2006vi,Buras:2004ub,London:2004uj,Fleischer:2007hj}. 
Experimental investigation of the issue is therefore very desirable, and has the potential to enable a significant 
advancement of the field. The \Bspipi\  and  \BdKK\ decay modes are ideal for this investigation, 
because all quarks in the final state are different from those in the initial state, so they can be 
mediated solely by amplitudes with penguin-annihilation ($\mathit{PA}$) and $W$-exchange ($E$) topologies (see Fig.~\ref{fig:diagrams}). 
However, they have not yet been observed, the best upper limits at 90\% CL being 
respectively $1.2\times 10^{-6}$ ~\cite{Aaltonen:2008hg} and $0.41\times 10^{-6}$~\cite{Abe:2006xs}.
A simultaneous measurement of branching fractions of 
both modes would be especially useful, 
as it would allow a better constraint on the strength of $\mathit{PA}$ and $E$ amplitudes~\cite{Buras:2004ub}.


In this Letter we report the results of a
simultaneous search for the two decays \Bspipi\ and \BdKK~\cite{C-conjugate},
using data corresponding to 6\lumifb\ integrated luminosity of $\bar{p}p$ collisions at $\sqrt{s} = 1.96$ TeV,
collected by the upgraded Collider Detector (CDF II) at the Fermilab Tevatron.  


The CDF II detector is described in detail in Ref.~\cite{Acosta:2004yw} with the detector subsystems relevant for this 
analysis discussed in Ref.~\cite{Abulencia:2006psa}.
The data are collected by a three-level online event-selection system (trigger).
At level~1, tracks are reconstructed 
in the transverse plane~\cite{CDF-coordinates}.
Two opposite-charge  particles are required, with reconstructed transverse momenta $p_{T1}, p_{T2} > 2~\pgev$, 
the scalar sum $p_{T1}+p_{T2} > 5.5~\pgev$, and an azimuthal opening angle $\Delta\phi < 135^{\circ}$. 
At level~2, tracks are combined with silicon-tracking-detector hits
and their impact parameter $d$ (transverse distance of closest approach to the beam line) 
is determined with 45 \micron\ resolution (including the beam spread) and required to be $0.1 < d < 1.0$ mm. 
A tighter opening-angle requirement, $20^{\circ} < \Delta\phi < 135^{\circ}$, is also applied.
Each track pair is then used to form a $B$ candidate, which is  
required to have an impact parameter $d_B < 140~\micron$ and to have traveled a 
distance $\Lxy > 200~\micron$ in the transverse plane. At level~3,  a cluster of computers confirms the selection with a full event reconstruction. 


The offline selection is based on a more accurate 
determination of the same quantities used in the trigger, with the 
addition of two further observables: 
the isolation ($I_{B}$) of the $B$ candidate~\cite{Isolation},
and the quality of the three-dimensional fit ($\chi^{2}$  with $1^{\circ}$ degree of freedom) of the 
decay vertex of the $B$ candidate. 
Requiring isolated candidates further reduces the background from
light-quark jets, and a low $\chi^{2}$ reduces the 
background from decays of different long-lived particles within the event, 
owing to the good resolution of the silicon-tracking detector in the $z$ direction.
We use the same final selection originally devised for the \BsKpi\ search~\cite{Aaltonen:2008hg}, 
whose simulation has proven to be nearly optimal also for detection of \Bspipi. 
This includes the following criteria: 
$I_B>0.525$, $\chi^{2}<5$, $d > 120~\micron$, $d_B < 60~\micron$, and $\Lxy > 350~\micron$.
%

\begin{figure}[tb]
\includegraphics[scale=0.35]{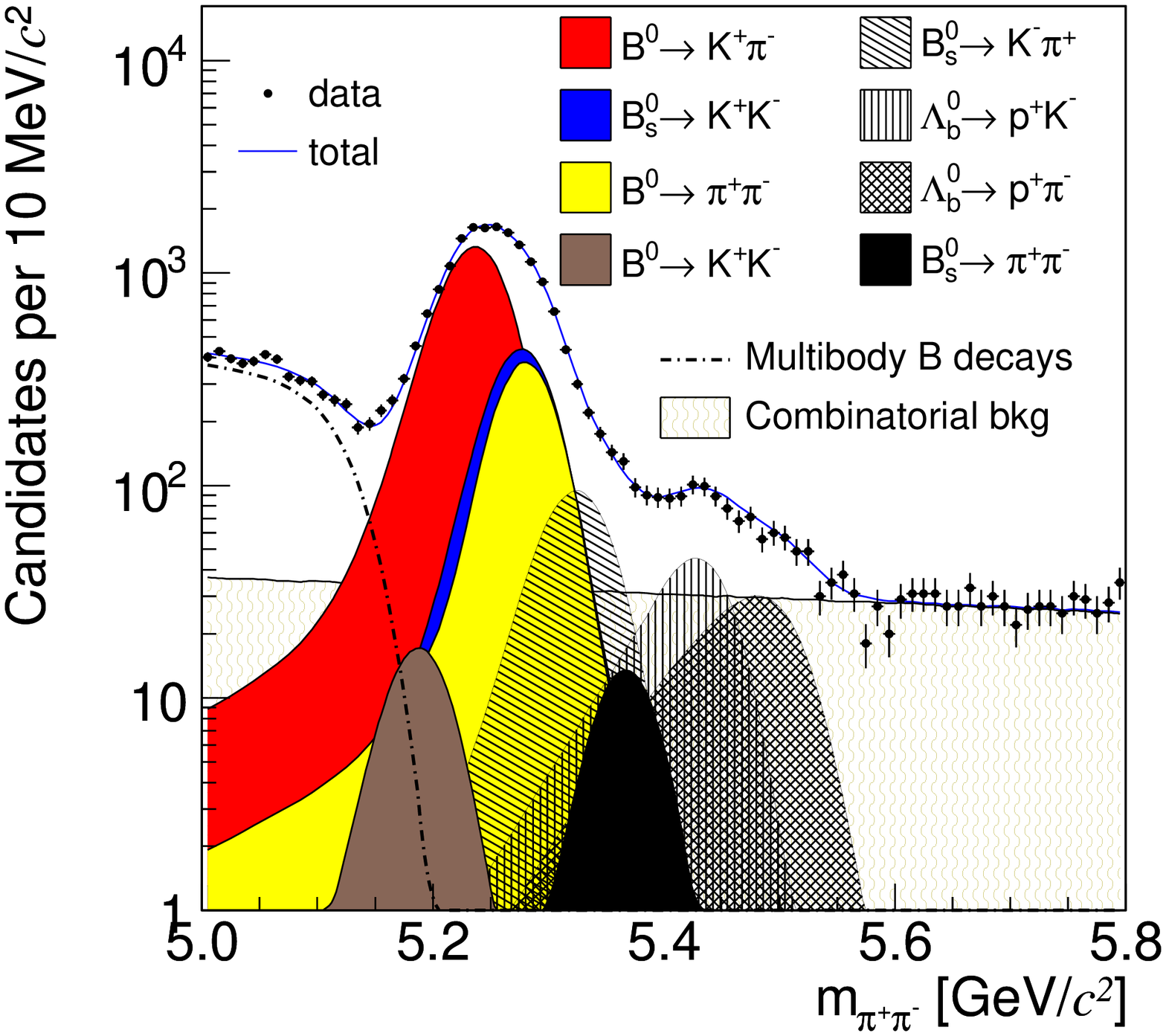}
\caption{\label{fig:projections} Mass distribution of reconstructed candidates. The charged pion mass is assigned to both tracks.
The sum of the fitted distributions and the individual components of signal and background are overlaid on the data distribution. 
}
\end{figure}

At most one $B$ candidate per event is found after this 
selection, and a mass ($m_{\pi^+\pi^-}$) is assigned to 
each, using a charged pion mass assignment for both decay products. 
The resulting mass distribution is shown in Fig.~\ref{fig:projections}, and is
dominated by the overlapping contributions of the \BdKpi, \Bdpipi, and \BsKK\ 
modes~\cite{Abulencia:2006psa,Aaltonen:2011qt}, with backgrounds coming from misreconstructed multibody $b$--hadron decays (physics background) and random pairs of charged particles (combinatorial background).
A \BdKK\ signal would appear in this distribution as an enhancement around $5.18~\massgev$,
while a \Bspipi\ signal is expected at the nominal $B^0_s$ mass of $5.3663~\massgev$,
where other more abundant modes also contribute~\cite{Aaltonen:2008hg}.


We used an extended unbinned likelihood fit, incorporating kinematic (kin) 
and particle-identification (PID) information, to determine the fraction of each individual mode in the sample.
The likelihood is defined as
\begin{equation}
 \mathcal{L} =  \frac{\nu^{N}}{N!}e^{-\nu} \cdot \prod^{N}_{i=1}  \mathcal{L}_i
\end{equation}
where $N$ is the total number of observed candidates, $\nu$ is the estimator of $N$ to be determined by the fit, 
 and  the likelihood for the $i$th event is
\begin{eqnarray}\label{eq:likelihood}
    \mathcal{L}_i & = & (1-b)\sum_{j} f_j \mathcal{L}^{\mathrm{kin}}_j  \mathcal{L}^{\mathrm{PID}}_j \nonumber \\
                  &   & +  b \left( f_{\rm{p}} \mathcal{L}^{\mathrm{kin}}_{\mathrm{p}}
    \mathcal{L}^{\mathrm{PID}}_{\mathrm{p}}+
   (1-f_{\rm{p}}) \mathcal{L}^{\mathrm{kin}}_{\mathrm{c}}
    \mathcal{L}^{\mathrm{PID}}_{\mathrm{c}}
	\right),
\end{eqnarray}
where the index $j$ runs over all signal modes, 
 and the index `p' (`c') labels the physics (combinatorial) 
background terms. The $f_j$ are the signal fractions to be determined by 
the fit, together with the background fraction parameters $b$ and $f_{\rm{p}}$.

For each charged hadron pair, the kinematic information is summarized by three loosely correlated
observables: the squared mass
$\mpipi^2$; the charged momentum asymmetry
$\beta = (p_{+}-p_{-})/(p_{+}+p_{-})$, where $p_+$ ($p_-$) is the
momentum of the positive (negative) particle; and  the scalar sum of particle momenta 
$\ptot=p_+ + p_-$.
The above variables allow evaluation of the squared invariant mass $m^{2}_{a^{+}b^{-}}$ of a candidate for 
any mass assignment of the positive and negative decay products ($m_{a^+}$,$m_{b^-}$), using the equation
\begin{eqnarray}\label{eq:Mpipi2}
 m^{2}_{a^+b^-}  =  m^{2}_{\pi^+\pi^-}  -  m_{\pi^+}^2  - m_{\pi^-}^2+ m_{a^+}^2+m_{b^-}^2 +
                 \nonumber    \\
          -  2 \sqrt{p_{+}^2+m_{\pi^+}^2} \sqrt{p_{-}^2+m_{\pi^-}^2}  
                    +  2\sqrt{p_{+}^2+m_{a^+}^2} \sqrt{p_{-}^2+m_{b^-}^2},
 \end{eqnarray}
 where
$p_+ = \ptot \frac{1+\beta}{2}$ , $p_- = \ptot \frac{1-\beta}{2}$. 

\begin{table}
\caption{\label{tab:yields} Yields and significances of rare mode signals.
The first quoted uncertainty is statistical; the second is systematic.} 

{
\begin{tabular}{lcc}
\hline \hline
Mode 	& N$_{s}$	&  Significance 			\\
\hline
\BdKK	& 120 $\pm$  49 $\pm$ 42	& $2.0\sigma$	\\
\Bspipi	& 94 $\pm$  28 $\pm$ 11	& $3.7\sigma$	\\
\hline \hline
\end{tabular}
}
\end{table}

The likelihood terms $\like^{\mathrm{kin}}_{j}$ describe the kinematic distributions of $\mpipi^2$, $\beta$, and \ptot\ variables
 for the physics signals and are obtained from Monte Carlo simulations.  
The same distributions for the combinatorial background are instead extracted from real data~\cite{Ruffini-Thesis}, and are inserted into the 
likelihood  through the $\like^{\mathrm{kin}}_{\mathrm{c}}$ term.
%
In particular, the squared-mass distribution of the combinatorial background is parametrized by an exponential function. 
The slope is fixed in the fit to the value extracted from an enriched sample of two
generic random tracks, containing events passing all requirements of final selections 
except for vertex quality, replaced by an antiselection cut  $\chi^2 > 40$, which strongly rejects track pairs originating from a common vertex.
The likelihood term $\like^{\mathrm{kin}}_{\mathrm{p}}$ describes the kinematic distributions 
of the background from partially reconstructed decays of generic $B$ hadrons. The $\mpipi^2$ distribution is, in this case,  modeled 
by an ARGUS function~\cite{Albrecht:1990am} convoluted with a Gaussian resolution,
while $\beta$ and \ptot\ distributions are obtained from Monte Carlo simulation. 

The fit has 28 free parameters. A detailed description of the fit and its parameters can be found in
Ref.~\cite{Morello-Thesis,Ruffini-Thesis}.

To ensure the reliability of the search for small signals in 
the vicinity of larger peaks, the shapes of the mass distributions assigned to each signal have been 
modeled in detail.  Momentum dependence and non--Gaussian resolution tails are
accounted for  by a full simulation of the detector, while the effects of soft photon 
radiation in the final state are simulated by {\sc photos}~\cite{photos}.
This resolution model was accurately checked against the observed shape 
of the $3.2\times 10^{6}$ \DKpi\ and  $140 \times 10^{3}$ \Dpipi\ signals in a sample of  $D^{*+}\to  D^0\pi^+$
 decays, collected with a similar trigger selection.  
As a result, the systematic uncertainty related to the signal mass shapes 
is negligible with respect to other uncertainties.

The $D^{*+}\to D^0\pi^+$ sample was also used to calibrate the \dedx\ response of the drift chamber 
to kaons and pions, using the charge of the $D^{*+}$ pion to identify the $D^0$ decay
products. The \dedx\ response of protons was determined from a sample of about 167~000 
$\Lambda\to p \pi^{-}$ decays, where the kinematic properties and the momentum threshold of the trigger allow unambiguous identification 
of the decay products~\cite{Morello-Thesis}. 
PID information is summarized by a single observable $\kappa$, defined as:
\begin{equation}
\kappa \equiv \frac{{  d}E/{  d}x - {  d}E/{d}x(\pi)}{{  d}E/{  d}x(K) - {  d}E/{  d}x(\pi)}
\end{equation}
where ${  d}E/{  d}x(\pi)$ and ${  d}E/{  d}x(K)$ are the expected ${ d}E/{  d}x$ depositions for those particle assignments.
The average values of $\kappa$ expected for pions and kaons are by construction 0 and 1.
Statistical separation between kaons and pions is about 1.4$\sigma$, while the 
ionization rates of protons and kaons are quite similar in the momentum range of interest. 
The PID likelihood term, which is similar for physics signals and backgrounds, 
depends only on $\kappa$ and on its expectation value $\langle \kappa \rangle$ (given a mass hypothesis) of the decay products.
In particular the physics signals model is described by the likelihood term $\mathcal{L}_{j}^{\mathrm{PID}}$, where the 
index $j$ uniquely identifies the final state,
while the background model is described 
by the two terms $\mathcal{L}_{\mathrm{p}}^{\mathrm{PID}}$ and $\mathcal{L}_{\mathrm{c}}^{\mathrm{PID}}$, 
respectively for the physics and combinatorial background, 
 that account for all possible pairs that can be formed combining only pions and kaons.
In fact muons are indistinguishable from pions  with the available \dedx\ resolution, and are therefore included within the nominal pion component.
For similar reasons, the small proton component in the background has been included within the nominal kaon component.
Thus the physics background model allows for independent, charge-averaged contributions 
of pions and kaons, whose fractions are determined by the fit;
while the combinatorial background model, instead, allows for more contributions, since independent fractions of 
 positively and negatively charged pions and kaons are determined by the fit. 


The signal fractions returned by the fit are in agreement with those obtained in the previous iteration of this analysis~\cite{Aaltonen:2008hg}.
The yields for the \Bspipi\ and \BdKK\ modes, obtained from those fractions, are shown in Table~\ref{tab:yields}.
The significance is evaluated as the ratio of the yield observed in
data to its total uncertainty (statistical and systematic uncertainties
added in quadrature), where the statistical uncertainty is determined from a simulation where the size of that signal is set to zero.
This evaluation assumes a Gaussian distribution of yield estimates, 
supported by the results obtained from repeated fits to simulated samples.
This procedure yields a more accurate measure of significance 
than the purely statistical estimate obtained from $\sqrt{-2\Delta {\rm ln}({\cal L})}$.

\begin{figure}[!tb]
\includegraphics[scale=0.35]{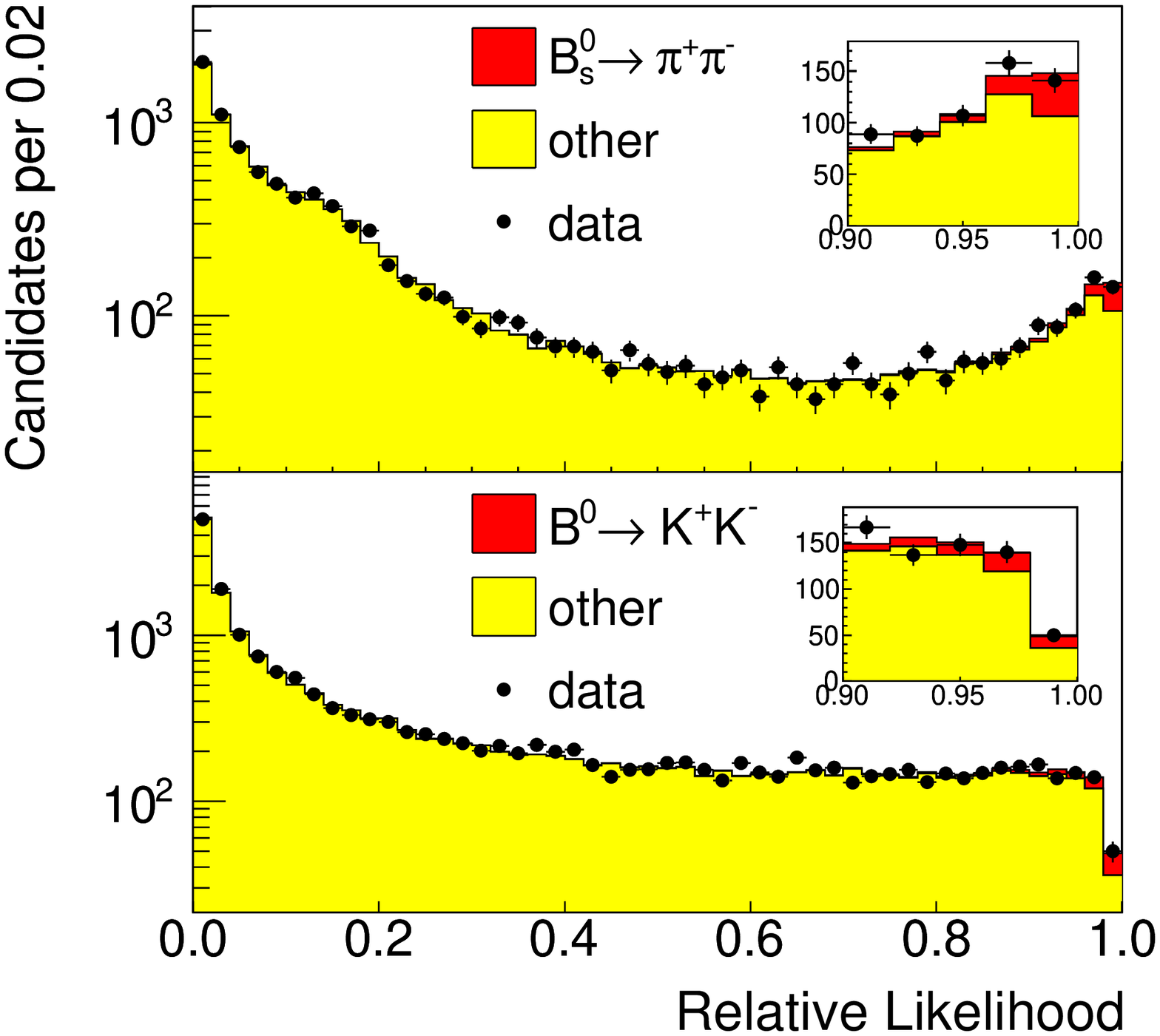}
\caption{Distribution of the relative signal likelihood, $\like_{S}/(\like_{S}+\like_{{\rm other}})$, in the region $5.25 <\mpipi<
5.50~\massgev$ for \Bspipi\ and $5.10 <\mpipi< 5.35~\massgev$ for \BdKK. For each event, $\like_{S}$ is the 
likelihood for the \Bspipi\ (top panel) and \BdKK\ (bottom panel) signal hypotheses, and $\like_{\rm other}$ is the likelihood for everything but the chosen signal, i.e., the weighted combination of all other 
components according to their measured fractions. Points with error 
bars show the distributions of data and histograms show the distributions predicted from the measured fractions. 
Zoom of the region of interest is shown in the inset.}
\label{fig:LRplots}
\end{figure}

We obtain a $3.7\sigma$  significant signal for the \Bspipi\ mode, and
we observe an excess at the $2.0\sigma$ level for the \BdKK\ mode.
As a check on the method, Fig.~\ref{fig:LRplots} shows relative likelihood distributions for these modes, which are in good agreement with our model.

\begin{table*}[htb]
\caption{\label{tab:BRs} Measured relative branching fractions of
rare modes. 
Absolute branching fractions were derived by normalizing to the current world--average value
${\mathcal B}(\mbox{\BdKpi}) = (19.4\pm 0.6) \times 10^{-6}$, and assuming the average values at high energy for the production fractions: 
$f_{s}/f_{d}= 0.282\pm 0.038$~\cite{PDG}.
The first quoted uncertainty is statistical; the second is systematic.}

{
\begin{tabular}{lcrclcccc}
\hline\hline
Mode && \multicolumn{3}{c}{Relative \BR}   && Absolute \BR\ ($10^{-6}$)&& Limit ($10^{-6}$)\\
\hline
\BdKK	&& \BdKKsuBdKpidef   &=&       0.012 $\pm$ 0.005 $\pm$ 
0.005          && 0.23 $\pm$ 0.10 $\pm$ 0.10 &&  $[0.05,0.46]$ at 90\% C.L.\\
\Bspipi	&& \BspipisuBdKpidef   &=&       0.008 $\pm$ 0.002 
$\pm$ 0.001        && 0.57 $\pm$ 0.15 $\pm$ 0.10 &&  $-$\\
\hline\hline
\end{tabular}
}
\end{table*}

As a further check an alternate fit was performed, using kinematic information only. Removal of \dedx\ information leads to results 
in agreement with the main fit, but with a loss in resolution of a factor 2 for \Bspipi\ and 3 for \BdKK , confirming the importance of this information. 

%

To avoid large uncertainties associated with production cross sections 
and absolute reconstruction efficiency, we measure all branching 
fractions relative to the \BdKpi\ mode.
A frequentist limit~\cite{Feldman:1997qc} at the 90\% C.L. is quoted for the 
\BdKK\ mode.
The raw fractions returned by the fit are corrected for the differences in selection efficiencies 
among different modes, which do not exceed 10\%. 
These corrections are determined from detailed detector simulation, 
with only two exceptions that are measured from data:
the momentum-averaged relative isolation efficiency between \Bs\ and \Bd, and 
the difference in efficiency for triggering on kaons and pions due to 
the different specific ionization in the drift chamber.
The former  is determined as $1.00 \pm 0.03$ from fully-reconstructed samples of
\Bs $\rightarrow J/\psi\phi$, and \Bd $\rightarrow J/\psi K^{*0}$ decays~\cite{Morello-Thesis}. 
The latter  is determined  from samples of $D^0$ mesons decaying into pairs of charged hadrons~\cite{Ruffini-Thesis}.
We measure the relative branching fractions 
$\mathcal{B}(D^0 \to \pi^{+}\pi^{-})/\mathcal{B}(D^0 \to K^{-}\pi^{+})$ and $\mathcal{B}(D^0 \to K^{+}K^{-})/\mathcal{B}(D^0 \to K^{-}\pi^{+})$.
The numbers of events are extracted from the available samples of tagged  $D^0 \to \pi^+\pi^-$, $D^0 \to K^{-}\pi^{+}$ and $D^0 \to K^{+}K^{-}$ decays, fitting the invariant $D^{*}\pi$ mass spectrum~\cite{Ruffini-Thesis},
while reconstruction efficiencies are determined from the same simulation used for the measurements described in this Letter.  
Comparison of these numbers with world measurement averages~\cite{PDG} allows us to extract the correction needed to compensate for the different efficiency of the tracking trigger for kaons and pions. 
The final corrections applied to our result do not exceed 5\% and are independent of particle momentum. 


The dominant contribution to the systematic uncertainty on both branching fractions is due to the \dedx\ model, which derives from the statistical uncertainty on
 the 48 parameters used for the analytical description of 
 the correlated \dedx\ response of the two decay products~\cite{Morello-Thesis}.  
This uncertainty is evaluated by repeating the likelihood fit 200 times with different sets of those parameters, randomly extracted from 
a multidimensional sphere, centered on the central value of the parametrization, with a radius 
corresponding to 1$\sigma$ of statistical uncertainty. 
The correlations between the parameters are neglected because 
their total effect, known from Ref.~\cite{Aaltonen:2011uu}, where they have been accounted for in detail,  
brings a reduction of the final systematic uncertainty because most correlations are negative. 
The \dedx-induced  systematic uncertainty on each observable is then obtained as the standard deviation of the distribution 
of that observable, over the ensemble of likelihood fits performed with different sets of parameters.
This approach is adequate for our purposes since the statistical uncertainty is greater than or of the same order of the systematic
uncertainty. 

The second dominant contribution to the systematic uncertainty for \Bspipi\ comes from the uncertainty on 
the relative efficiency correction, while for \BdKK\ it comes from the uncertainty in the background model, 
which includes a sizeable component of partially reconstructed decays with poorly known branching fractions. 
The latter systematic uncertainty is conservatively assessed by performing extreme variations of the assumed 
relative contributions of the various modes in the simulation; the resulting uncertainty is still a factor of 2 
lower than the uncertainty associated to the \dedx\ model.

Other contributions come from trigger efficiencies, 
$b$--hadron masses, $b$--hadron lifetimes and $\Delta\Gamma_s/\Gamma_s$, and transverse momentum distribution of the \Lb\ baryon.
A further systematic uncertainty of the order of 10\% is included
for the \BdKK\ mode to account for a small bias of the fitting procedure observed in simulated samples.


The final results are listed in Table \ref{tab:BRs}. Absolute 
branching fractions are also quoted, by normalizing to world-average 
values of production fractions and 
\BR(\BdKpi)~\cite{PDG}. 
The branching fraction measured for the \Bspipi\ mode is consistent 
with and supersedes the previous upper limit ($< 1.2 \times 10^{-6}$ at 90\%~C.L.), based on a subsample of the current data~\cite{Aaltonen:2008hg}. 
It is in agreement with predictions obtained with the pQCD approach~\cite{Ali:2007ff,Li:2004ep}, but it is higher than most 
other theoretical predictions \cite{Beneke:2003zv,Sun:2002rn,Cheng:2009mu}.
The central value for $\BR(\BdKK)$ is the most precise determination of this quantity to date, 
and is in agreement with previous experimental results~\cite{Abe:2006xs,Aubert:2006fha} and theoretical predictions~\cite{Beneke:2003zv,Cheng:2009mu}.
It supersedes the previous CDF limit~\cite{Aaltonen:2008hg}, based on a subsample of the current data.
The present measurements represent a significant step  
in reducing a source of uncertainty in many theoretical predictions for charmless $B$-decays.
The results favor a large annihilation scenario, which 
is somewhat unexpected for instance in QCDF~\cite{Zhu:2011mm}.  


In summary, we have searched in CDF data for as-yet-unmeasured charmless decay modes
of neutral $b$~mesons into pairs of charged mesons.
We report an updated upper limit for the \BdKK\ mode and the first evidence for the \Bspipi\ mode and a measurement of its branching fraction.

\begin{acknowledgments}
We thank the Fermilab staff and the technical staffs of the participating 
institutions for their vital contributions. This work was supported by 
the U.S. Department of Energy and National Science Foundation; the 
Italian Istituto Nazionale di Fisica Nucleare; the Ministry of Education, 
Culture, Sports, Science and Technology of Japan; the Natural Sciences 
and Engineering Research Council of Canada; the National Science Council 
of the Republic of China; the Swiss National Science Foundation; the A.P. 
Sloan Foundation; the Bundesministerium f\"ur Bildung und Forschung, 
Germany; the Korean World Class University Program, the National Research Foundation
of Korea; the Science and Technology Facilities Council and the Royal 
Society, UK; 
the Russian Foundation for Basic Research; the
Ministerio de Ciencia e Innovaci\'{o}n, and Programa Consolider-Ingenio 2010, Spain;
the Slovak R\&D Agency; the Academy of Finland; and the Australian Research
Council (ARC).

\end{acknowledgments}
 %

\begin{thebibliography}{37}
\bibitem[{CDF()}]{CDF-coordinates}
\emph{\bibinfo{title}{{\em CDF II uses a cylindrical coordinate system in which
 $\phi$ is the azimuthal angle, $r$ is the radius from the nominal beam line,
 and $z$ points in the proton beam direction, with the origin at the center of
 the detector. The transverse plane is the plane perpendicular to the $z$
 axis.}}}
\end{thebibliography}
 %



\end{document}